\documentstyle[aps,preprint]{revtex}
\newcommand{\eps}{\epsilon}
\newcommand{\pdt}{\partial_t}
\newcommand{\pdx}{\partial_x}

\newcommand{\pdy}{\partial_y}

\newcommand{\dxi}{\partial_\xi}
\newcommand{\deta}{\partial_\eta}
\newcommand{\dtau}{\partial_\tau}
\newcommand{\domega}{\dot\omega}
\newcommand{\ddomega}{\ddot\omega}

\begin{document}
\title{Derivation of Amplitude Equations by Renormalization Group
 Method }
\author{Ken-ichi Matsuba and Kazuhiro Nozaki\\
Department of Physics,Nagoya University,Nagoya 464-01,Japan}
\maketitle
\begin{abstract}
A proper formulation in the perturbative renormalization group method 
is presented to deduce amplitude equations. 
 The formulation makes it possible not only
 avoiding a serious difficulty in the previous reduction to
amplitude equations by eliminating all of the secular terms but also
consistent derivation of higher-order correction to amplitude equations.
\end{abstract}
Recently a novel method based on the perturbative renormalization group
 (RG) theory has been developed as an asymptotic singular 
perturbation technique by L.Y.Chen, N.Goldenfeld and Y.Oono \cite{oono1}.
Their renormalization group method ( RGM) removes 
secular or divergent terms from a perturbation series 
by renormalizing integral constants of  lower order solutions. 
Although they have impressively succeeded  in application to 
ordinary differential equations  \cite{oono2}, a consistent formulation
of the RG method does not have been established for application 
to partial differential equations. In fact, we have recently pointed
out \cite{rmatsu},\cite{rcmatsu} that all of secular solutions of 
a perturbation series does not have  
been eliminated yet in the previous derivation of amplitude equations 
from partial differential equations \cite{oono2},\cite{graham} . \par
 In this paper, we present a suitable formulation of 
 a perturbation problem in RGM in order to avoid such a difficulty
 and deduce some amplitude equations and their higher-order correction
 consistently in the framework of RGM.
 A crucial part of our formulation is to scale all variables except one
 independent variable so that the non-scaled variable is taken as 
 a parameter of RG. Therefore, the renormalization group equations (RGE)
 or amplitude equations depend on the choice of scale, that is, the initial
 setting of a perturbation problem, of which significance is pointed out
 by S.Sasa \cite{sasa}. \par
(I) As the first example, we derive an 
amplitude equation from the following weakly nonlinear wave equation:
\begin{equation}
\pdt^2u-\pdx^2u+(1+\eps^2u^2)u=0,\label{nwe}
\end{equation}
where $\eps$ is a small parameter. In order to focus our attention to
 a slowly-varing amplitude, let us introduce a complex amplitude $A$
 and a stretched variable $\xi$:
 \begin{equation}
u=A\exp[i(kx-\omega t)]+c.c.,\quad\xi=\eps x,\label{str}
\end{equation}
where $k$ is a wavenumber, $\omega=\sqrt{1+k^2}$  and  $c.c.$ denotes 
complex conjugate, then Eq.(\ref{nwe}) is rewritten as
\begin{equation}
[LA-\eps(2ik\dxi+\eps\dxi^2)A+3\eps^2|A|^2A]e^{i\theta}
+\eps^2A^3e^{3i\theta}+c.c.=0,\label{amp1}
\end{equation}
where $L\equiv\pdt^2-2i\omega\pdt$ and $\theta=kx-\omega t$.
In this formulation, $t$ is chosen as a parameter of RG.   
Substituting an expansion
\begin{equation}
A=A_0(\xi)+\eps A_1+\eps^2 A_2+\eps^3 A_3+\cdots,\label{expn}
\end{equation}
 into Eq.(\ref{amp1}) and retaining only secular contribution,
  we have,up to order $\eps^3$,
\begin{eqnarray*} 
LA_1&=&2ik\dxi A_0,\\
LA_2&=&2ik\dxi A_1+\dxi^2A_0-3|A_0|^2A_0,\\
LA_3&=&2ik\dxi A_2+\dxi^2A_1-6|A_0|^2A_1-3A_0^2\bar A_1,
\end{eqnarray*}
where $\bar A$ is complex conjugate of $A$. 
Noting that secular (polynomial) solutions of $LP_j=t^j$ are given by
$$P_0=\frac{it}{2\omega},\quad P_1=\frac{it^2}{4\omega}+\frac{t}{4\omega^2},
\quad P_2=\frac{it^3}{6\omega}+\frac{t^2}{4\omega^2}
-\frac{it}{4\omega^3},$$
we obtain the following secular solution up to order $\eps^3$
\begin{eqnarray}
A&=&A_0+t[-\eps\domega\dxi A_0+\eps^2\frac{i}{2}(\ddomega\dxi^2A_0-
\frac{3}{\omega}|A_0|^2A_0)
 \nonumber\\
&-&\eps^3\frac{\domega}{2\omega}\dxi
(\ddomega\dxi^2A_0-\frac{3}{\omega}|A_0|^2A_0)] 
+\frac{t^2}{2}(\eps^2\domega)[\domega\dxi A_0
\nonumber \\
&-&i\eps\dxi(\ddomega\dxi^2A_0-\frac{3}{\omega}|A_0|^2A_0)]
-\eps^3\frac{t^3}{6}\domega^3\dxi^3A_0,\label{sc1}
\end{eqnarray}
where $\domega=d\omega/dk$ and $\ddomega=d\domega/dk$.
All of secular terms in Eq.(\ref{sc1}) are removed by renormalizing $A_0$
and the renormalized amplitude $A$ is described by RGE: 
\begin{eqnarray}
\pdt A&=&-\eps\domega\dxi A+\eps^2\frac{i}{2}(\ddomega\dxi^2A-
\frac{3}{\omega}|A|^2A)-\frac{\domega}{2\omega}\dxi
(\ddomega\dxi^2A-\frac{3}{\omega}|A|^2A),\label{ns1}\\
\pdt^2 A&=&\eps^2\domega[\domega\dxi A-i\eps\dxi
(\ddomega\dxi^2A-\frac{3}{\omega}|A|^2A),\label{ns2}\\ 
\pdt^3 A&=&-\eps^3\domega^3\dxi^3A,\label{ns3}
\end{eqnarray}
where Eq.(\ref{ns1}) is the nonlinear Schr\"odinger equation with
 correction up to order $\eps^3$, while Eqs.(\ref{ns2}) and
(\ref{ns3}) are easily shown to be compatible with Eq.(\ref{ns1}) 
up to order $\eps^3$. Since the above procedure  continues to
arbitrary order in $\eps$, we can derive systematically 
the nonlinear Schr\"odinger equation with correction up to arbitrary order.
\par 
  
(II) Let us derive  slowly-varing amplitude equations from 
the Swift-Hohenberg equation \cite{swft}:  
\begin{equation}
[\pdt+(1+\triangle)^2-\eps^2(1-u^2)]u=0,\label{swf}
\end{equation}
where $\triangle\equiv\pdx^2+\pdy^2$. Introducing a complex amplitude  
$A$ and stretched variables $\tau$ and $\eta$ as
$$u=A\exp(ix)+c.c.,\quad\tau=\eps^m t,\quad\eta=\eps^n y,$$
where $m$ and $n$ are positive constants,
 we rewrite Eq.(\ref{swf})  as
\begin{equation}
[MA+2\eps^{2n}(2i\pdx+\pdx^2)\deta^2A+\eps^{4n}\deta^4A
+\eps^m \dtau A-\eps^2(1-3|A|^2)A]e^{ix}
+\eps^2A^3e^{3ix}+c.c.=0,\label{amp2}
\end{equation}
where $M\equiv(\pdx^2+2i\pdx)^2$ and $x$ is  a parameter of 
renormalization group. 
When we set $m=2$ and $n=1/2$, we obtain the secular solution of 
Eq.(\ref{amp2}),through the perturbative procedure similar to 
the example (I), as
\begin{eqnarray}
A&=&A_0(\tau,\eta)+x\eps C_0(\tau,\eta)
\nonumber\\
&+&\frac{x^2}{2}[-\frac{\eps^2}{4}B_0
+\frac{\eps^3}{4}(\frac{1}{2}\deta^2 B_0+i(\dtau+\deta^4-1+6|A_0|^2)C_0
 +3iA_0^2\bar C_0]
 \nonumber\\
 &-&\frac{x^3}{6}\frac{\eps^3}{4}[i\deta^2 B_0-
(\dtau+\deta^4-1+6|A_0|^2)C_0-3A_0^2\bar C_0],\label{sc2}
\end{eqnarray}
where $\bar C$ is complex conjugate of $C$ and
$$B_0=-(\dtau+\deta^4-1+3|A_0|^2)A_0-4i\deta^2 C_0.$$
Renormalizing $A_0\to A,\quad B_0\to B$ and $C_0\to C$ so that secular 
terms in  Eq.(\ref{sc2}) are removed, we get the following RGE.  
\begin{eqnarray}
\pdx A&=&\eps C,\label{nw1}\\
\pdx^2 A&=&-\frac{\eps^2}{4}B+\frac{\eps^3}{4}[\frac{1}{2}\deta^2 B
+i(\dtau+\deta^4-1+6|A|^2)C
+3iA^2\bar C],\label{nw2}\\
\pdx^3 A&=&-\frac{\eps^3}{4}[i\deta^2 B-(\dtau+\deta^4-1+6|A|^2)C
-3A^2\bar C].\label{nw3}
\end{eqnarray}
Eliminating $C$ from Eqs.(\ref{nw1}) and (\ref{nw2}), we have 
the Newell-Whitehead-Segel equation \cite{nw}
 with correction up to order $\eps^3$.
$$\pdx^2 A=i\eps(1-\frac{\eps}{2}\deta^2)\pdx\deta^2 A+\frac{\eps^2}{4}
(1+i\pdx-\frac{\eps}{2}\deta^2)(\dtau+\deta^4-1+3|A|^2)A.$$
It is easy to see that Eq.(\ref{nw3}) is compatible with Eqs.(\ref{nw1})
 and (\ref{nw2}).
 If we set $m=2$ and $n=1$, the similar procedure yields another
 amplitude equation up to order $\eps^3$ :
 $$\pdx^2 A=\frac{\eps^2}{4}(1+i\pdx)(\dtau-1+3|A|^2)A+
 i\eps^2\pdx\deta^2A.$$
 (III)As a final example, we derive a slow amplitude equation from
the following model equation:
\begin{equation}
\{\pdt[\pdt+(k^2+\triangle)^2-\eps^2(1-u^2)]-\triangle\}u
+(1+\eps^2u^2)u=0,\label{hpb}
\end{equation}
The nonlinear wave equation (\ref{nwe}) and 
the Swift-Hohenberg equation (\ref{swf}) are combined in 
the model equation (\ref{hpb}) ,which 
may be one of the simplest equations describing the Hopf
bifurcation in continuous media. 
Let us choose $t$ as a parameter of RG and introduce a  complex
amplitude $A$ and a stretched variable $\xi$ in the same way as the
 first example (see Eq.(\ref{str})) and $\eta=\eps y$,
 then we can derive the following two-dimensional complex Ginzburg-Landau 
 equation \cite{matsu} with correction up to order $\eps^3$  
 after following the similar procedure used in the above examples:
 $$(\pdt+\eps\domega\dxi) A=\frac{i\eps^2}{2}B-\frac{\domega\eps^3}{2\omega}
 \dxi(\ddomega\dxi^2+\frac{\domega}{k}\deta^2-3|A|^2)A,$$
 where
 $$B=[(\ddomega-4ik^2)\dxi^2+\frac{\domega}{k}\deta^2-i-3(1-i)|A|^2]A.$$
\par
 In summary, we present a proper formulation in the perturbative 
 renormalization group method and deduce typical amplitude equations and
 their higher-order correction consistently. In our formulation, we can 
 avoid the serious difficulty appearing in the previous derivation
 of amplitude equations  by changing the initial
  setting of a perturbation problem.

\end{document}